# NUCLEAR HYDRODYNAMICS IN THE INNER CRUST OF NEUTRON STARS*

Piotr Magierski[1] and Aurel Bulgac[2]

[1]Faculty of Physics, Warsaw University of Technology
ul. Koszykowa 75, 00-662 Warsaw, Poland
Piotr.Magierski@olimp.if.pw.edu.pl
[2]Department of Physics, University of Washington,
Seattle, WA 98195-1560, USA
bulgac@phys.washington.edu

In the inner crust of a neutron star, due to the high density and pressure, nuclei which are still present, are immersed in a neutron superfluid. One then expects that the dynamical properties of nuclei are significantly affected. In order to estimate the magnitude of the effect associated with the presence of a superfluid medium, we formulate the hydrodynamical approach to the nuclear dynamics in the inner crust of neutron stars. We calculate the renormalized nuclear mass and the strength of the medium-induced interaction between nuclei. We argue that these effects noticeably modify the properties of the Coulomb crystal in the inner crust.

## 1. Introduction

The inner crust of neutron stars has a complex structure which depends on the nucleon density. In the outer parts of the inner crust spherical nuclei are immersed in a neutron superfluid. They form a crystal lattice stabilized by the Coulomb interaction. At larger densities the phase transition to the deformed nuclear phase occurs, leading eventually to the predicted existence of a great variety of exotic nuclear structures. The static properties of the nuclear matter under such extreme conditions have been a subject of a considerable theoretical effort [1, 2, 3, 4, 5, 6, 7, 8, 9, 10, 11, 12, 13, 14, 15,

---

16]. In particular, the phase transition pattern at the bottom of the crust has been analyzed using methods based on the density functional theory.

However, in order to analyze the thermal and electric conductivities of the crust, which are important for understanding of the thermal evolution of neutron stars one has to go beyond the static approximation. Since the part of the inner crust is filled with nuclei likely forming a Coulomb crystal, the dynamical properties of such systems have been studied by many authors (see e.g. [17] and references therein). An important ingredient of the theory of a Coulomb crystal is the plasma frequency, which defines the energy scale for the lattice vibrations. It depends on the ion density, ion charge and masses. Although the two first quantities can be calculated using the usual density functional approach, the last one requires some caution, since the nuclei are immersed in a fermionic environment and their bare masses are renormalized, in order to take into account the interaction with the surrounding neutrons. Namely, as nuclei oscillate around the stable positions they bring the fermionic medium into motion as well.

The example we are considering here is a particular realization of the problem of a heavy particle (impurity) dynamics in a Fermi system. In general, one may expect that the equation of motion for such an impurity moving through the fermionic medium could be expressed in the form:

$$M^{ren}\frac{d^2q}{dt^2} + \eta\frac{dq}{dt} + \frac{dE_{int}}{dq} = 0, \tag{1}$$

where $q$ is a coordinate describing the position of the nucleus, $\eta$ is the dissipation coefficient and $E_{int}$ denotes the reversible part of interaction energy with surrounding fermions. There are two problems associated with the above equation. First, one has to determine the mass $M^{ren}$ of the impurity immersed in the fermionic medium. Certainly a bare mass of an impurity should be renormalized, since the impurity travels in the fermionic medium in a cloud of medium excitations. The second problem is associated with the fact that the simple Newtonian equation of motion with friction cannot be translated directly to quantum mechanics. There is no analogue of the concept of friction in quantum mechanics as friction is not time-reversal invariant. The solution to this puzzle is the coupling of the considered impurity to an environment, which consist of various excitations of the Fermi liquid. However deriving from this approach an effective equation of motion for the impurity is usually based on some serious approximations. There are a number of papers on the problem [18, 19, 20, 21, 22, 23, 24, 25, 26, 27, 28], which however have little overlap with the present approach. The reason is threefold, first, we consider the rather particular case of an impurity (nucleus) which is not only heavy, but also large (i.e. $k_F R \gg 1$, where $k_F$ is the Fermi momentum and $R$ is a nuclear radius). Second, the neutron bath in which the nucleus travels is incompressible, and last but not least, the system is superfluid, since the temperatures we are interested in, are far below the critical temperature for loss of superfluidity, which is believed to be of the order of 1 MeV.

Hence in the following we will make certain assumptions dictated by the special properties of the nuclear system. Namely, we assume that the neutrons outside nuclei form an incompressible and irrotational liquid. It implies that their dynamics can be described by a velocity potential. Second, we neglect the energy dissipation effects, assuming that the energy transfer between the nucleus and the medium is a fully time-reversible process. This assumption seems to be well justified for sufficiently low temperatures. Although it is clear that at finite temperatures there always exists a gas of quasiparticle excitations, which are responsible for energy dissipation, however if we limit our analysis to small temperatures, as compared to the critical temperature of the system, we may disregard this effect, since the number of quasiparticle excitations decreases exponentially with decreasing temperature. Last, but not least, we will assume that the flow velocity induced by a moving nuclear impurity is below the critical velocity for the loss of superfluidity. Again it is justified for low energy lattice excitations which are the most interesting in the context of the neutron star crust.

## 2. Hydrodynamical approach to the nuclear dynamics in a superfluid fermionic environment

### 2.1. Single nucleus in a superfluid medium

Let us consider a spherical nucleus of a constant density $\rho_{in}$ and of radius $R$ in the fluid at the density $\rho_{out}$. The nuclear matter, both inside and outside a nucleus is assumed to be incompressible and irrotational, which implies that there exist a velocity potential $\Phi(\vec{r}) = \Phi_{in}(\vec{r}) + \Phi_{out}(\vec{r})$, where $\Phi_{in}(\vec{r})$ and $\Phi_{out}(\vec{r})$ stand for the velocity field inside and outside a nucleus, respectively. As a consequence of the continuity requirement, the function $\Phi$ fulfills the Poisson equation:

$$\nabla^2 \Phi_{out} = 0; \qquad \text{for } r > R, \qquad (2)$$
$$\nabla^2 \Phi_{in} = 0; \qquad \text{for } r < R. \qquad (3)$$

The velocity field $\vec{v}(\vec{r}) = \vec{v}_{in}(\vec{r}) + \vec{v}_{out}(\vec{r})$ is related to the velocity potential through the well known equations:

$$\vec{v}_{in} = \nabla \Phi_{in}; \qquad \text{for } r > R, \qquad (4)$$
$$\vec{v}_{out} = \nabla \Phi_{out}; \qquad \text{for } r < R. \qquad (5)$$

The boundary conditions are determined by the geometry of the system. In the case of a spherical nucleus moving with the velocity $\vec{u}$ the following conditions have to be imposed on $\Phi$:

$$\begin{aligned}\Phi_{in}|_{r=R} &= \Phi_{out}|_{r=R}, \\ \rho_{in}\left(\frac{\partial}{\partial r}\Phi_{in} - \vec{n}\cdot\vec{u}\right)\Big|_{r=R} &= \rho_{out}\left(\frac{\partial}{\partial r}\Phi_{out} - \vec{n}\cdot\vec{u}\right)\Big|_{r=R}, \\ \Phi_{out}|_{r\to\infty} &= 0,\end{aligned} \quad (6)$$

The vector $\vec{n}$ is the outward normal to the surface of a spherical nucleus and $r$ is a radial coordinate (the center of a nucleus is placed at $r = 0$). The first equation stems from the requirement that the phase of the wave function of the superfluid system is continuous, the second one expresses a conservation of mass in radial flow, and finally the third one ensures that the correct asymptotic behavior for the velocity field is obtained.

The solution of eq. (2) with the boundary conditions (6) leads to the following expressions for the velocity potential:

$$\Phi_{in}(\vec{r}) = \frac{1-\gamma}{2\gamma+1}\vec{u}\cdot\vec{r} \quad (7)$$

$$\Phi_{out}(\vec{r}) = \frac{1-\gamma}{2\gamma+1}\left(\frac{R}{r}\right)^3\vec{u}\cdot\vec{r}, \quad (8)$$

where $\gamma = \dfrac{\rho_{out}}{\rho_{in}}$. Note that in the limit of $\gamma \to \infty$, which correspond to the case of an empty bubble moving through the fluid, one recovers the well known result [29, 30].

Thus $\Phi$ describes a dipolar flow of the liquid around the impurity. The kinetic energy of the liquid associated with the flow reads:

$$\begin{aligned}T &= \frac{1}{2}m\int_V \rho|\nabla\Phi|^2 d^3r = \\ &= \frac{1}{2}m\rho_{in}\oint_S \Phi_{in}\frac{\partial\Phi_{in}}{\partial r}dS - \frac{1}{2}m\rho_{out}\oint_S \Phi_{out}\frac{\partial\Phi_{out}}{\partial r}dS,\end{aligned} \quad (9)$$

where $m$ is the nucleon mass and $S$ is the surface of the nucleus. The last term appeared as a consequence of the Green's theorem. Applying the boundary conditions one obtains:

$$T = \frac{1}{2}m(\rho_{in} - \rho_{out})\oint_S \Phi_{in}|_{r=R}\vec{u}\cdot\vec{n}dS. \quad (10)$$

Hence the kinetic energy reads:

$$T = \frac{2}{3}\pi R^3 m\rho_{in}\frac{(1-\gamma)^2}{2\gamma+1}u^2, \quad (11)$$

and it is clear that the effective mass of the bubble is equal to

$$M^{ren} = \frac{4}{3}m\rho_{in}\pi R^3 \frac{(1-\gamma)^2}{2\gamma+1} = M\frac{(1-\gamma)^2}{2\gamma+1}, \qquad (12)$$

where $M$ denotes the bare mass of the nucleus.

Note that the effective mass depends on the nucleon density. It is instructive to consider various limiting cases of the above formula, namely:

$$\begin{aligned}
M^{ren} &= \frac{2}{3}m\rho_{out}\pi R^3 = \frac{\gamma}{2}M & \text{for } \gamma \to \infty \text{ (bubble)}, & \qquad (13) \\
M^{ren} &= 0 & \text{for } \gamma = 1 \text{ (uniform system)}, & \qquad (14) \\
M^{ren} &= \frac{4}{3}m\rho_{in}\pi R^3 = M & \text{for } \gamma = 0 \text{ (nucleus in a vacuum)}. & \qquad (15)
\end{aligned}$$

The above expressions show that the nuclear mass for a nucleus immersed in a superfluid neutron liquid is decreased as compared to the bare mass $M$ (since $\rho_{in} > \rho_{out}$ in the case of a nucleus in a low density neutron fluid). Moreover, the renormalized mass depends on the density of the surrounding neutrons. This suggests then that dynamical properties of the Coulomb crystal will be strongly affected especially in the deeper layers of the inner crust. Indeed, the plasma frequency in the inner crust changes as compared to the value calculated for the case: $\rho_{out} = 0$ (ion in the vacuum). Namely, $\frac{\omega_p}{\omega_p^{ren}} = \frac{|1-\gamma|}{\sqrt{2\gamma+1}}$, where $\omega_p^{ren}$ corresponds to the plasma frequency for the system with renormalized ion masses [31].

*2.2. Medium-induced interaction of two moving nuclei in a superfluid liquid*

Although the mass renormalization effect discussed in the previous subsection seems to influence the dynamics of the ion crystal in the inner crust of neutron star, the picture is still not complete without taking into account the medium-induced interaction with other ions. Namely when the single nucleus is moving through the neutron liquid it produces a neutron flow, which affects the motion of neighbouring ions as well. One may expect that due to the fact that the potential flow generated by a moving nucleus behaves like $1/r^2$, it may induce a long range correlations in the system, and would supplement the standard Coulomb interaction between ions.

Let us consider two nuclei moving with velocities $\vec{u}_i$ ($i = 1, 2$). The nuclei have radii $R_i$ and are placed at the distance $r$ (see Fig. 1). The velocity potential can be expressed in the form:

$$\Phi = \sum_{i=1}^{2}(\Phi_{i\ in} + \Phi_{i\ out}) = \sum_{i=1}^{2}\Phi_{i\ in} + \Phi_{out}, \qquad (16)$$

$$\Phi_{out} = \sum_{i=1}^{2} \Phi_{i\,out}, \tag{17}$$

where $\Phi_{i\,in}$ is the velocity potential inside the $i$−th nucleus and $\Phi_{i\,out}$ is the outside potential depending only on the velocity $\vec{u}_i$. The boundary conditions which have to be fulfilled in this case read:

$$\begin{aligned}
\Phi_{i\,in}|_{r=R_i} &= \Phi_{out}|_{r=R_i}, \\
\rho_{in}\left(\frac{\partial}{\partial r}\Phi_{i\,in} - \vec{n}_i \cdot \vec{u}_i\right)\bigg|_{r=R_i} &= \rho_{out}\left(\frac{\partial}{\partial r}\Phi_{out} - \vec{n}_i \cdot \vec{u}_i\right)\bigg|_{r=R_i}, \\
\Phi_{out}|_{r\to\infty} &= 0; \qquad \text{for i=1,2,}
\end{aligned} \tag{18}$$

where $\vec{n}_i$ is the outward normal to the surface of the $i$-th nucleus. We have also assumed here that the density inside each nucleus is the same and equal to $\rho_{in}$. It is a justified assumption although it is easy to generalize the following expressions for the case of two different nuclear densities.

Fig. 1. Two nuclei of radii and velocities $R_1, \vec{u}_1$ and $R_2, \vec{u}_2$, respectively

The problem can be solved perturbatively, using the method of images, provided the distance between the nuclei is large. Namely, if $r = |\vec{r}_1 - \vec{r}_2|$ is large as compared to $R_1$ and $R_2$ then the natural small parameter appearing in the expansion of the velocity potential is $\dfrac{R_1 R_2}{r^2}$. Up to the second order in the expansion, the velocity potential outside nuclei can be expressed in the form:

$$\begin{aligned}
\Phi_{1\,out} &= A_1 \frac{\vec{r}_1 \cdot \vec{u}_1}{r_1^3} \\
&+ \frac{1}{2}(A_1 + \delta A_1)\left(\frac{R_2}{r}\right)^3 \left(\frac{\vec{r}_2 \cdot \vec{u}_1}{r_2^3} - \frac{3}{r^2}\frac{(\vec{r} \cdot \vec{u}_1)(\vec{r}_2 \cdot \vec{r})}{r_2^3}\right)
\end{aligned}$$

$$+ O\left(\frac{1}{r^6}\right) \tag{19}$$

$$\Phi_{2\,out} = A_2 \frac{\vec{r}_2 \cdot \vec{u}_2}{r_2^3}$$
$$+ \frac{1}{2}(A_2 + \delta A_2)\left(\frac{R_1}{r}\right)^3 \left(\frac{\vec{r}_1 \cdot \vec{u}_2}{r_1^3} - \frac{3}{r^2}\frac{(\vec{r}\cdot\vec{u}_2)(\vec{r}_1\cdot\vec{r})}{r_1^3}\right)$$
$$+ O\left(\frac{1}{r^6}\right). \tag{20}$$

It is easy to check that, with the above choice of the outside velocity field, the following limiting values of the potential and its radial derivatives are obtained:

$$\Phi_{out}\mid_{r_1=R_1} = A_1 \frac{\vec{R}_1 \cdot \vec{u}_1}{R_1^3} - A_2 \frac{\vec{r}\cdot\vec{u}_2}{r^3}$$
$$+ \left(\frac{3}{2}A_2 + \frac{1}{2}\delta A_2\right)\left(\frac{R_1}{r}\right)^3 \left(\frac{\vec{R}_1 \cdot \vec{u}_2}{R_1^3} - \frac{3}{r^2}\frac{(\vec{r}\cdot\vec{u}_2)(\vec{R}_1\cdot\vec{r})}{R_1^3}\right)$$
$$+ O\left(\frac{1}{r^4}\right) \tag{21}$$

$$\Phi_{out}\mid_{r_2=R_2} = A_2 \frac{\vec{R}_2 \cdot \vec{u}_2}{R_2^3} + A_1 \frac{\vec{r}\cdot\vec{u}_1}{r^3}$$
$$+ \left(\frac{3}{2}A_1 + \frac{1}{2}\delta A_1\right)\left(\frac{R_2}{r}\right)^3 \left(\frac{\vec{R}_2 \cdot \vec{u}_1}{R_2^3} - \frac{3}{r^2}\frac{(\vec{r}\cdot\vec{u}_1)(\vec{R}_2\cdot\vec{r})}{R_2^3}\right)$$
$$+ O\left(\frac{1}{r^4}\right) \tag{22}$$

$$\frac{\partial}{\partial r_1}\Phi_{out}\mid_{r_1=R_1} = -2A_1 \frac{\vec{R}_1 \cdot \vec{u}_1}{R_1^4}$$
$$- \delta A_2 \left(\frac{R_1}{r}\right)^3 \left(\frac{\vec{R}_1 \cdot \vec{u}_2}{R_1^4} - \frac{3}{r^2}\frac{(\vec{r}\cdot\vec{u}_2)(\vec{R}_1\cdot\vec{r})}{R_1^4}\right)$$
$$+ O\left(\frac{1}{r^4}\right) \tag{23}$$

$$\frac{\partial}{\partial r_2}\Phi_{out}\mid_{r_2=R_2} = -2A_2 \frac{\vec{R}_2 \cdot \vec{u}_2}{R_2^4}$$

$$- \delta A_1 \left(\frac{R_2}{r}\right)^3 \left(\frac{\vec{R}_2 \cdot \vec{u}_1}{R_1^4} - \frac{3}{r^2}\frac{(\vec{r} \cdot \vec{u}_1)(\vec{R}_2 \cdot \vec{r})}{R_2^4}\right)$$
$$+ O\left(\frac{1}{r^4}\right), \tag{24}$$

where $\vec{R}_i = R_i \frac{\vec{r}_i}{r_i}$. Hence the inside solution can be written as:

$$\begin{aligned}
\Phi_{1\,in} &= A_1 \frac{\vec{r}_1 \cdot \vec{u}_1}{R_1^3} - A_2 \frac{\vec{r} \cdot \vec{u}_2}{r^3} \\
&+ \left(\frac{3}{2}A_2 + \frac{1}{2}\delta A_2\right)\left(\frac{R_1}{r}\right)^3 \left(\frac{\vec{r}_1 \cdot \vec{u}_2}{R_1^3} - \frac{3}{r^2}\frac{(\vec{r} \cdot \vec{u}_2)(\vec{r}_1 \cdot \vec{r})}{R_1^3}\right) \\
&+ O\left(\frac{1}{r^4}\right),
\end{aligned} \tag{25}$$

$$\begin{aligned}
\Phi_{2\,in} &= A_2 \frac{\vec{r}_2 \cdot \vec{u}_2}{R_2^3} + A_1 \frac{\vec{r} \cdot \vec{u}_1}{r^3} \\
&+ \left(\frac{3}{2}A_1 + \frac{1}{2}\delta A_1\right)\left(\frac{R_2}{r}\right)^3 \left(\frac{\vec{r}_2 \cdot \vec{u}_1}{R_2^3} - \frac{3}{r^2}\frac{(\vec{r} \cdot \vec{u}_1)(\vec{r}_2 \cdot \vec{r})}{R_2^3}\right) \\
&+ O\left(\frac{1}{r^4}\right); \qquad \text{for } r_1 < R_1 \text{ and } r_2 < R_2.
\end{aligned} \tag{26}$$

The above form of the inside solutions ensure that the first pair of boundary conditions (18) is automatically fulfilled. The other boundary conditions, originating from the conservation of the mass flow, determine the unknown coefficients $A_i, \delta A_i$, namely:

$$A_i = R_i^3 \frac{\gamma - 1}{2\gamma + 1}, \tag{27}$$

$$\delta A_i = -3R_i^3 \frac{\gamma - 1}{(2\gamma + 1)^2}. \tag{28}$$

The total kinetic energy of the system consisting of two nuclei immersed in the superfluid medium reads:

$$\begin{aligned}
T &= \frac{1}{2}m\rho_{in}\left(\int_{r_1 < R_1}|\nabla\Phi_{1\,in}|^2 d^3r_1 + \int_{r_2 < R_2}|\nabla\Phi_{2\,in}|^2 d^3r_2\right) \\
&+ \frac{1}{2}m\rho_{out}\int_{V_{out}}|\nabla\Phi_{out}|^2 d^3r_1,
\end{aligned} \tag{29}$$

where $V_{out} = V - \frac{4}{3}\pi(R_1^3 + R_2^3)$ is the volume ocuppied by the outside neutrons. Analogously as in the case of a single nucleus one can transform the above equation using the Green's theorem and the boundary conditions:

$$T = \frac{1}{2}m(\rho_{in} - \rho_{out}) \oint_{S_1} \Phi_{1\,in}|_{r=R_1}(\vec{u}_1 \cdot \vec{n}_1)dS_1$$
$$+ \frac{1}{2}m(\rho_{in} - \rho_{out}) \oint_{S_2} \Phi_{2\,in}|_{r=R_2}(\vec{u}_2 \cdot \vec{n}_2)dS_2. \qquad (30)$$

Hence finally we get:

$$T = \frac{1}{2}(M_1^{ren}u_1^2 + M_2^{ren}u_2^2)$$
$$+ 4\pi m \rho_{out} \left(\frac{1-\gamma}{2\gamma+1}\right)^2 \left(\frac{R_1 R_2}{r}\right)^3 \left[\vec{u}_1 \cdot \vec{u}_2 - \frac{3}{r^2}(\vec{u}_1 \cdot \vec{r})(\vec{u}_2 \cdot \vec{r})\right] \quad (31)$$

where the renormalized masses of nuclei have the form:

$$M_i^{ren} = \frac{4}{3}m\rho_{in}\pi R_i^3 \frac{(1-\gamma)^2}{2\gamma+1} = M_i \frac{(1-\gamma)^2}{2\gamma+1}, \qquad (32)$$

for $i = 1, 2$ and $M_i$ denotes the nuclear bare mass of the $i-th$ nucleus.

As one can notice besides the usual kinetic energy of the two nuclei there appears also a crossterm, which in the case of two identical nuclei has the form:

$$E_{int} = 4\pi m \rho_{out} \left(\frac{1-\gamma}{2\gamma+1}\right)^2 \left(\frac{R^2}{r}\right)^3 \left[\vec{u}_1 \cdot \vec{u}_2 - \frac{3}{r^2}(\vec{u}_1 \cdot \vec{r})(\vec{u}_2 \cdot \vec{r})\right] =$$
$$= 3M^{ren}\frac{\gamma}{2\gamma+1}\left(\frac{R}{r}\right)^3 \left[\vec{u}_1 \cdot \vec{u}_2 - \frac{3}{r^2}(\vec{u}_1 \cdot \vec{r})(\vec{u}_2 \cdot \vec{r})\right], \qquad (33)$$

giving rise to the mutual interaction which is a manifestation of the Bernoulli law. The velocity dependent interaction between two nuclei has the same $1/r^3$ behavior as the (static) Fermionic Casimir interaction between two nuclei in a neutron Fermi sea, see Refs. [10, 11, 12, 13, 14].

One may raise the following question: Does the medium-induced coupling between neighbouring nuclei (33) represent a significant correction to the Coulomb interaction which stabilize the crystal? The problem is apparently innocuous since the interaction energy (33) depends on the velocity of nuclei, whereas the Coulomb energy is a position dependent function. Nevertheless, it is important to know whether the contributions coming from the hydrodynamic coupling represent only a small correction to the dynamical matrix, or may significantly modify the properties of the crystal.

In the harmonic approximation one can estimate the ratio of the averaged hydrodynamic interaction energy and the Coulomb interaction energy, associated with the thermal excitations [32]. It turns out that the leading term is proportional to $\frac{\gamma}{2\gamma+1}\frac{R}{R_c}$, where $R_c$ is the radius of the Wigner-Seitz cell. Hence the hydrodynamical coupling turns out to be important in the bottom layers of the inner crust.

## 3. Conclusion

In the present paper we have formulated the hydrodynamic approach to the nuclear dynamics in the inner crust of neutron stars. We have derived the formula for the renormalized nuclear masses, assuming that the nuclei are immersed in an incompressible neutron superfluid. We argued that the renormalization of the mass affects the phonon spectrum of the crystal in a significant manner. Moreover, we have calculated the velocity dependent induced interaction between two moving nuclear impurities, which is a manifestation of the Bernoulli law. This interaction becomes important inside the inner crust.

Summarizing, our investigations indicate that the dynamical properties of the nuclei in the neutron star inner crust are noticeably modified by the presence of superfluid neutron medium. One may expect that besides the translational degrees of freedom of the ions, also the nuclear shape vibrations can significantly influence the dynamic of the Coulomb crystal. Clearly, various nuclear collective degrees of freedom may turn out to be important for understanding of thermal and electric conductivities of the crust.

## Acknowledgements

PM was supported in part by the Polish Committee for Scientific Research (KBN) under Contract No. 5 P03B 014 21 and AB by US DOE. One of the authors (PM) would like to thank the Nuclear Theory Group for hosting his visit in Seattle. We thank Andreas Wirzba and Karlheinz Langanke for discussions.